\documentclass{article}
\usepackage[utf8]{inputenc}
\usepackage[english]{babel}
\usepackage{graphicx}
\usepackage{hyperref} 
\usepackage{geometry}
\usepackage{amsmath}
\usepackage{amssymb}
\usepackage{amsfonts}
\usepackage{multicol}
\usepackage{float}
\usepackage{caption}
\usepackage{subcaption}
\usepackage{cite}
\usepackage{amsthm}
\usepackage[utf8]{inputenc}
\usepackage{xcolor}
\title{\bf Bounds on Very Weakly Interacting \\ Ultra Light Scalar  and Pseudoscalar Dark Matter \\ from Quantum Gravity}
\author{Xavier Calmet\thanks{E-mail: X.Calmet@sussex.ac.uk}$\ $ 
and Folkert~Kuipers\thanks{E-mail: F.Kuipers@sussex.ac.uk} 
\\
\\
{\em Department of Physics and Astronomy, University of Sussex,}\\ 
{\em Brighton, BN1 9QH, United Kingdom}
}

\begin{document}
\maketitle
%
\begin{abstract}
In this paper we consider very weakly interacting and ultra light scalar and pseudoscalar dark matter candidates. We show that quantum gravity has important implications for such models and that the masses of the singlet scalar and pseudoscalar fields must be heavier than $3\times 10^{-3}$ eV. However, if they are gauged, their masses could be much lighter and as light as $10^{-22}\, {\rm eV}$. The existence of new gauge forces in the dark matter sector can thus be probed by atomic clocks or quantum sensors experiments.
\end{abstract}
\flushbottom
%
\thispagestyle{empty}
\pagebreak
\pagenumbering{arabic}
\section{Introduction}

A strong evidence for physics beyond the standard model of particle physics comes from the observation that 75\% of the matter balance of our universe cannot be accounted for by the standard model. Some form of non-luminous matter must exist. 
Despite being the most abundant form of matter, embarrassingly little is known about dark matter and a wide range of masses and couplings to the standard model particles are still possible. In this paper, we focus on ultralight and very weakly coupled scalar and pseudoscalar dark matter models which have recently received a fair share of attention and for which a large part of the parameter space can now be probed experimentally \cite{Weinberg:1977ma,Preskill:1982cy,Hu:2000ke,Piazza:2010ye,Khmelnitsky:2013lxt,Porayko:2014rfa,Schive:2014dra,Beyer:2014uja,Marsh:2015xka,Urena-Lopez:2015gur,Calabrese:2016hmp,Blas:2016ddr,Bernal:2016lll,Bernal:2017oih,Hui:2016ltb}.

In particular experiments that search for oscillations in the fundamental constants resulting from the coupling of scalar or pseudoscalar dark matter with the standard model \cite{Arvanitaki:2014faa,Stadnik:2014tta,Stadnik:2015kia,Graham:2015ifn,Arvanitaki:2015iga,Stadnik:2015xbn,Stadnik:2016zkf,Abel:2017rtm,Smorra:2019qfx} have a great potential of testing such models in the mass range $m_\phi\in[10^{-16},10^{-23}]{\rm eV}$. The optimal sensitivity of such experiments typically lies around $10^{-22} \, {\rm eV}$, and the bounds on the sensitivity are set by the fact that the oscillation frequency is proportional to the mass of the scalar field. Masses of the order $m_\phi\sim10^{-16}\,{\rm eV}$ correspond to oscillation times of the order $T\sim10\,{\rm s}$, while masses of the order $m_\phi\sim10^{-23}\,{\rm eV}$ correspond to oscillation times of the order $T\sim10\,{\rm yr}$.

In this paper we follow the line of arguments put forward in refs. \cite{Calmet:2019jyz,Calmet:2019frv} based on quantum gravity to put further theoretical bounds on such searches. In particular, we exploit the fact that dark matter will always couple gravitationally to the standard model. Therefore quantum gravity will generate effective interactions between the standard model and the hidden sector. This fact together with current experimental bounds restricts the mass range for such weakly interacting light particles considerably. While this is the case for singlet scalar fields, we show that this is not the case if there are new forces in the dark matter sector.

\section{Interactions Generated by Quantum Gravity}
For any dark matter model we can write the following effective action.
\begin{equation}
	S = S_{\rm EH} + \int \sqrt{|g|}\left( \mathcal{L}_{\rm SM} + \mathcal{L}_{\rm DM} + \mathcal{L}_{\rm int}\right) d^4 x,
\end{equation}
where the standard model Lagrangian and the dark matter sector Lagrangian can be written as
\begin{align}
	\mathcal{L}_{\rm SM} &= \sum_i c_i \, \mathcal{O}_{{\rm SM},i},\\
	\mathcal{L}_{\rm DM} &= \sum_j c_j \, \mathcal{O}_{{\rm DM},j},
\end{align}
where $c_i,c_j$ are dimensionless Wilson coefficients. Interactions between the standard model particles and those of the dark matter section can be introduced via a Lagrangian
\begin{align}
	\mathcal{L}_{\rm int} &= \sum_k c_k \, \mathcal{O}_{{\rm int},k},
\end{align}
where again $c_k$ are dimensionless Wilson coefficients. 

Besides the ``particle physics" interactions induced by the operator $\mathcal{O}_{\rm int,k} $, there will be some gravitational interaction between the two sectors. Indeed, since both the standard model and the hidden sector couple to gravity, gravity will generate operators connecting the two sectors whether there is an interaction operator $\mathcal{O}_{\rm int,k}$ at tree level or not. 

For every $O_{{\rm SM},i}$ and $O_{{\rm DM},j}$, perturbative quantum gravity will generate the additional interactions $M_{\rm P}^{-4} O_{{\rm SM},i}\,O_{{\rm DM},j}$. We thus have 
\begin{equation}\label{PertInt}
	\mathcal{L}_{\rm int} = \sum_k c_k \, \mathcal{O}_{\rm int,k} 
	+ \sum_{i,j} \frac{c_{i,j}}{M_{\rm P}^4} \, \mathcal{O}_{{\rm SM},i}\,\mathcal{O}_{{\rm DM},j},
\end{equation}
where $M_{\rm P}$ is the reduced Planck scale, which is the scale of quantum gravity and where $c_{i,j}$ are Wilson coefficients of order unity.  It is clear from eq.~\eqref{PertInt} that the interactions generated by perturbative quantum gravity are suppressed by the reduced Planck scale to the fourth power. Therefore these interactions are not expected to be measurable in any contemporary or near future experiment. Hence, perturbative quantum gravity cannot yet provide any constraints to dark matter models.

Non-perturbative quantum gravity, on the other hand, can constrain dark matter models. Using the same argument, namely that everything couples to gravity as it is universal, one can deduce that non-perturbative quantum gravity effects could generate effective operators of any dimension. However any such operator must be suppressed by the scale of quantum gravity as such interactions must vanish in the limit where $M_{\rm P}\to \infty$, i.e. when gravity decouples. We thus expect quantum gravity induced effective interactions to be of the form
\begin{equation}
	\sum_{n \geq 0} \sum_{k} \tilde{c}_{n,k}\, \mathcal{O}_{{\rm QG},n,k} 
	= \sum_{n \geq 0} \sum_{k} \frac{\tilde{c}_{n,k}}{M_{\rm P}^n} \, O_{{\rm QG}, n,k},
\end{equation}
where $\mathcal{O}_{{\rm QG},n}$ has mass-dimension $4$ and $O_{{\rm QG}, n}$ has mass-dimension $n+4$.

As the Wilson coefficients $\tilde{c}_{d,k}$ depend on the ultra-violet completion of quantum gravity, one might be inclined to conclude that no predictions can be made until such a theory is known. However, experience with effective field theories, see discussion in  \cite{Calmet:2019jyz,Calmet:2019frv}, shows that sensible predictions on the order of magnitude of the Wilson coefficients can be made. Quite generically, Wilson coefficients are expected to be of order one, if the scale of the physics generating the interaction is known and properly normalized. In particular, there is no reason to expect an exponential suppression as it is sometimes claimed.  For example, it has been shown that there is no exponential suppression in the production of  quantum black holes in high energy collisions of particles  \cite{Hsu:2002bd}. 

In the case of quantum gravity, it is known that the scale of quantum gravity is dynamical. Naively, one might expect that the scale is the reduced Planck scale $M_{\rm P}=2.435\times 10^{18}$ GeV. However it is now well understood that the scale at which quantum gravitational interactions become relevant is $M_{\rm P} \sqrt{160 \pi/N}$ with $N=1/3 N_S+N_F+4 N_V$ where $N_S$, $N_F$ and $N_V$ are respectively the number of real scalar fields, Weyl fermions and vector bosons in the model \cite{Calmet:2014gya,Han:2004wt,Atkins:2010eq,Calmet:2008tn}. For the standard model, this is very close to the naive reduced Planck scale. Once the suppression scale for these operators has been properly defined there is no reason to expect a further suppression via smaller than unity Wilson coefficients. Furthermore, as we are considering non-perturbative physics, the Wilson coefficients will not be suppressed by loop factors or small coupling constants to some power.
Note that the scale of quantum gravity cannot be larger than the reduced Planck scale as adding more fields to the theory can only lead to a lower scale of quantum gravity. We are thus being as conservative as possible by taking the scale of quantum gravity to be the reduced Planck scale. 

We can now combine the quantum gravitational effective interactions with the non-gravitational interactions between the standard model and the dark matter sector. These can be written as
\begin{equation}
	\sum_k c_k \, \mathcal{O}_{{\rm int},k} 
	= \sum_{n \geq 0} \sum_{k} \frac{c_{n,k}}{\Lambda_{n,k}^n} \, O_{{\rm int},n,k},
\end{equation}
where $\Lambda_{n,k}$ is the energy scale associated with this effective operator. Comparing these two we find that non-gravitationally induced effective operators between the standard model and the hidden sector are corrected by gravitationally induced operators. Therefore, excluding all operators of dimension less than $4$, we can write down an interaction Lagrangian of the form
\begin{align}
	\mathcal{L}_{\rm int} 
	&=  \sum_{n \geq 0} \sum_{k} \left( \frac{c_{n,k}}{\Lambda_{n,k}^n} + \frac{\tilde{c}_{n,k}}{M_{\rm P}^n} \right)\, O_{{\rm int},n,k}\nonumber\\
	&=  \sum_{n \geq 0} \sum_{k} \frac{c_{n,k}}{\Lambda_{n,k}^n}\, 
	\left[ 1 + \frac{\tilde{c}_{n,k}}{c_{n,k}}\, 
	\left( \frac{\Lambda_{n,k}}{M_{\rm P}}\right)^n \right]\,O_{{\rm int},n,k}.
\end{align}
As both $\tilde{c}_{n,k}$ and $c_{n,k}$ are expected to be of order $1$, we find that the quantum gravitational interactions dominate, if $\Lambda_{n,k}>\, M_{\rm P}$. Note that $c_{n,k}$ could contain further loop suppression factors if the corresponding operators are generated perturbatively, but this does not change our analysis, the important point is that as we are considering nonperturbative quantum gravitational effects, there are no loop suppression factors in $\tilde{c}_{n,k}$.

Experiments looking for weakly interacting dark matter put bounds on the interaction strength $c_{n,k}/ \Lambda_{n,k}^n$. For some operators with $n\leq2$ these bounds have reached the Planck scale, i.e. $c_{n,k} M_{\rm P}^n \gtrsim \, \Lambda_{n,k}^n$. Therefore, since  $c_{n,k},\tilde{c}_{n,k}=\mathcal{O}(1)$, it is possible to exclude various models without probing more feeble interactions. In particular, if one operator can be excluded up to the Planck scale for a certain mass range, quantum gravity will exclude the existence of the scalar or pseudoscalar field for this mass range. This follows from the fact that quantum gravity will generate all possible, i.e. allowed by gauge symmetries, operators at the Planck scale.

\section{Scalar and Pseudoscalar Dark Matter}
In this section we discuss the consequences of the argument from the previous section for some specific scalar and pseudoscalar dark matter models. The most relevant models involving spinless dark matter are dimension $4$ operators. However, it is expected that the Wilson coefficients of dimension four operators must be exponentially suppressed by a factor $e^{-M_{\rm P}/\mu}$, as such quantum gravity induced operators should vanish in the limit $M_{\rm P}\rightarrow \infty$, i.e., when gravity decouples. Here $\mu$ is a renormalization scale.

The next most relevant operators for a spinless dark matter boson coupling to the standard model are dimension 5 operators. An example is an operator of the form
\begin{equation}\label{LinCoup}
	O_1 = \frac{c_1}{\Lambda_1} \, \phi \, F_{\mu\nu}F^{\mu\nu},
\end{equation}
where $\phi$ is the scalar dark matter field, and $F_{\mu\nu}$ is the electromagnetic field tensor. The results from the E\"{o}t-Wash torsion pendulum experiment that searches for fifth forces \cite{Kapner:2006si,Hoyle:2004cw,Adelberger:2006dh,Leefer:2016xfu,Braginsky1972,Smith:1999cr,Schlamminger:2007ht,Adelberger:2009zz,Zhou:2015pna} lead to the following bound\footnote{Bounds in the E\"ot-Wash experiments are usually presented in terms of the coupling strength $\alpha$ and the length scale of the Yukawa interaction $\lambda$. Such bounds can be translated into a mass-bound using the fact that $\alpha=\mathcal{O}(1)$ as discussed before and by noticing that $m_\phi c^2=\frac{\hbar c}{\lambda}$.}
\begin{equation}
	\frac{c_1}{\Lambda_1} \lesssim M_{\rm P}^{-1}
	\qquad {\rm if} \qquad 
	m_\phi \lesssim 3 \cdot 10^{-3}\, {\rm eV}
\end{equation}
and slightly stronger bounds for lower masses. Moreover atomic spectroscopy measurements \cite{VanTilburg:2015oza,Hees:2016gop} put even tighter bounds on such an interaction for masses $m_\phi \lesssim 10^{-18}\, {\rm eV}$, however these bounds rely on the assumption that the scalar field is the unique component of the dark matter sector.

\begin{figure}[h]
	\centering
	\includegraphics[width=8cm]{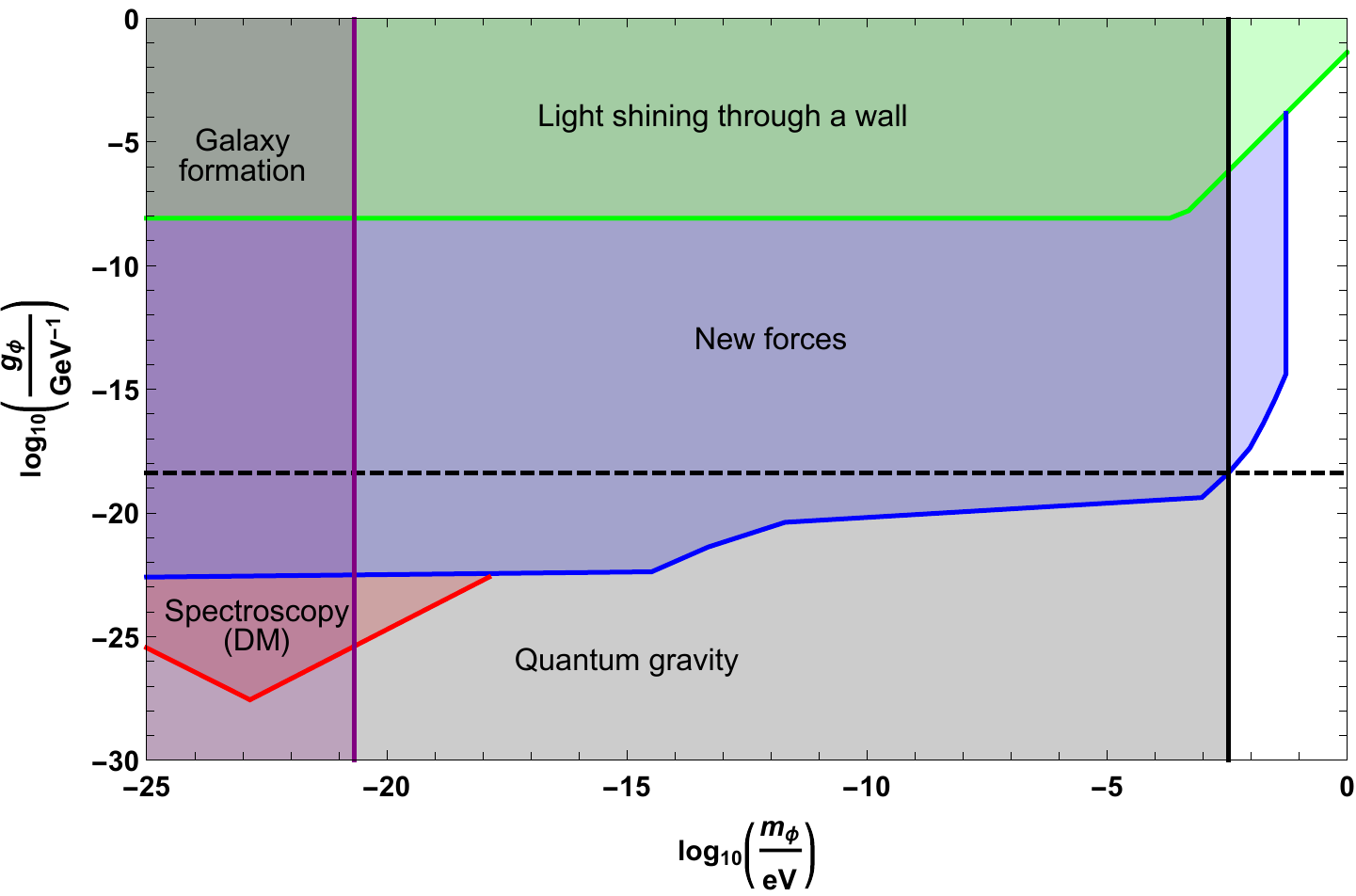}
	\caption{Limits on the linear scalar interaction $g_\phi=c_1/\Lambda_1$ as a function of the mass of the scalar $m_\phi$. Green: limits from light shining through a wall experiments \cite{Ehret:2010mh,Ballou:2015cka}. Blue: limits from torsion experiments \cite{Braginsky1972,Smith:1999cr,Hoyle:2004cw,Kapner:2006si,Adelberger:2006dh,Schlamminger:2007ht,Adelberger:2009zz,Zhou:2015pna,Leefer:2016xfu}. Red: limits from atomic spectroscopy experiments \cite{VanTilburg:2015oza,Hees:2016gop,Stadnik:2018sas}. Purple: limits from galaxy formation, quasar lensing  and stellar streams \cite{Bozek:2014uqa,Schive:2015kza,Corasaniti:2016epp,Irsic:2017yje,Nori:2018pka,Schutz:2020jox}. Black: limits from quantum gravity as discussed in this paper. Dashed black line: reduced Planck scale.}
	\label{PhiPlot}
\end{figure}

As argued above, quantum gravity will lead to an additional contribution
\begin{equation}\label{LinCoupQG}
	O_{1,{\rm QG}} = \left(\frac{c_1}{\Lambda_1}+\frac{\tilde c_1}{M_{\rm P}}\right) \, \phi \, F_{\mu\nu}F^{\mu\nu},
\end{equation}
with $\tilde c_1\sim {\cal O}(1)$ as argued before. Therefore the current bounds exclude this interaction for all masses $m_\phi\leq3 \times 10^{-3}{\rm eV}$. The resulting bounds on this interaction are summarized in figure \ref{PhiPlot}, which can be compared\footnote{Note that there is a factor $4$ difference: $g_{\phi}=\frac{g_\gamma^s}{4}$, where $g_{\phi}$ is the dimensionful coupling in this paper, and $g_\gamma^s$ is the dimensionful coupling in ref.~\cite{Stadnik:2018sas}.} to figure 31.1 in ref.~\cite{Stadnik:2018sas}.

\begin{figure}[h]
	\centering
	\includegraphics[width=8cm]{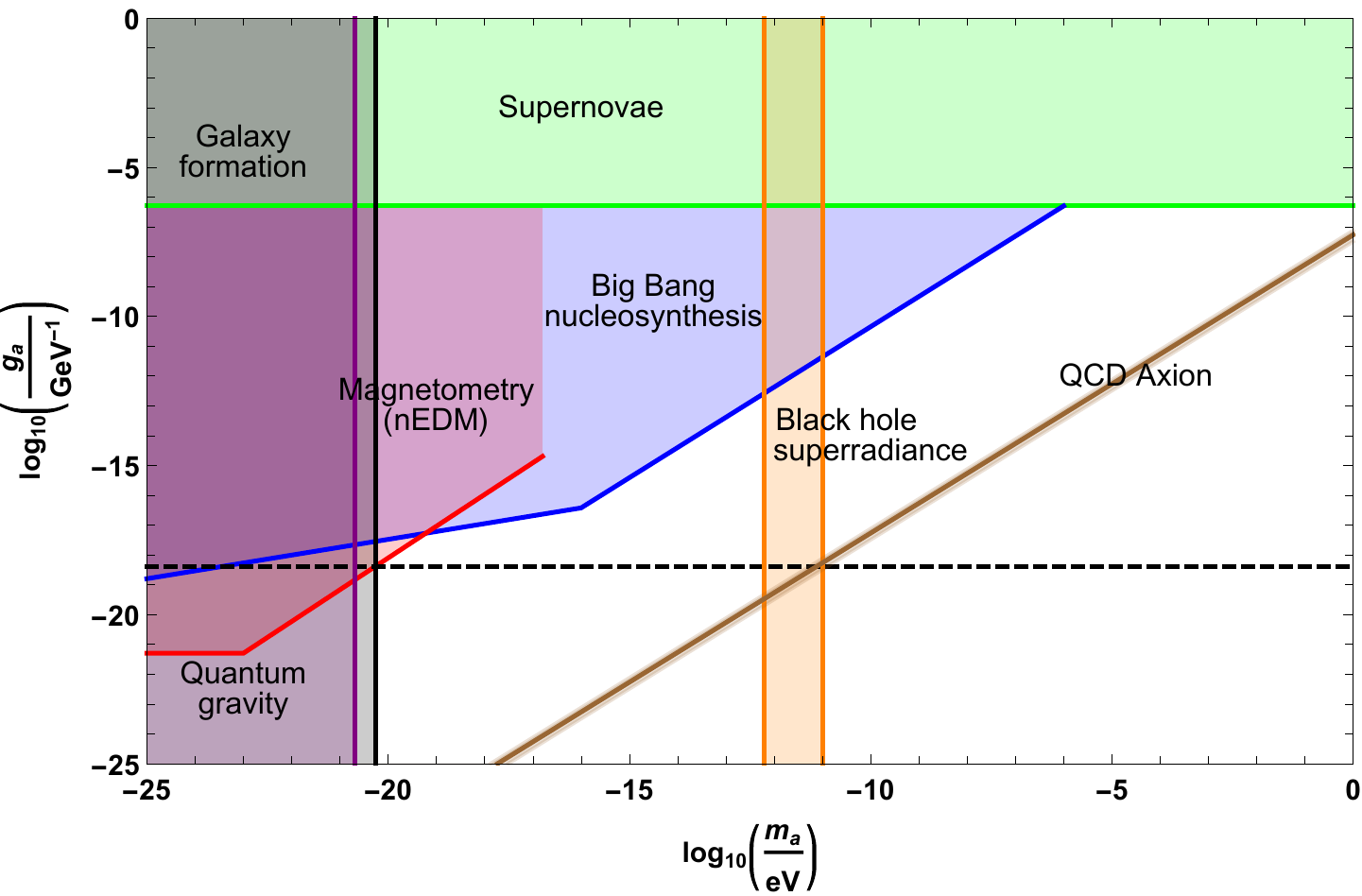}
	\caption{Parity conserving quantum gravity. Limits on the linear axion interaction $g_a=c_3/\Lambda_3$ as a function of the mass of the axion $m_a$. Green: limits from supernovae measurements \cite{Graham:2013gfa}. Blue: limits from the big bang nucleosynthesis \cite{Blum:2014vsa,Stadnik:2015uka,Stadnik:2015kia,Stadnik:2015upa,Stadnik:2017mid}. Red: limits from magnetometry experiments \cite{Abel:2017rtm,Stadnik:2018sas}. Purple: limits from galaxy formation, quasar lensing  and stellar streams\cite{Bozek:2014uqa,Schive:2015kza,Corasaniti:2016epp,Irsic:2017yje,Nori:2018pka,Schutz:2020jox}. Orange: limits from the superradiance instability of black holes \cite{Cardoso:2018tly}, however note that these bounds can be avoided, if the self-interaction of the axion-like particle is sufficiently strong \cite{Arvanitaki:2014wva}. Brown: predicted value of the QCD axion \cite{Pospelov:1999mv,Tanabashi:2018oca}. Black: axion masses below $m_a\lesssim 10^{-21} \, {\rm eV}$ are excluded by parity conserving quantum gravity as discussed in this paper. Dashed black line: reduced Planck scale.}
	\label{AxionPlot}
\end{figure}

Moreover, since quantum gravity generates interactions between all the particles of the standard model and the scalar field. Any scalar field with a mass below $3 \times 10^{-3}{\rm eV}$ would generate a Planck scale gravitational operator, which has not been detected by the E\"ot-Wash experiment. Therefore the derived bound does not exclusively apply to models containing the non-gravitationally induced interaction \eqref{LinCoup}. In fact, any dark matter model containing scalar dark matter fields of masses $m_\phi\lesssim 3 \times 10^{-3}\, {\rm eV}$ is excluded.
A similar analysis can be done for a pseudoscalar field $a$. The interaction between an axion-like-particle $a$ and gluons will receive a quantum gravitational correction
\begin{equation}\label{AxionPar}
	O_{2,{\rm QG}} = \left ( \frac{c_2}{\Lambda_2}+\frac{\tilde c_2}{M_{\rm P}} \right) \, a\, G_{\mu\nu}\tilde G^{\mu\nu},
\end{equation}
where $\tilde c_2\sim {\cal O}(1)$ and $G_{\mu\nu}$ is the usual gluonic field strength and $\tilde{G}^{\mu\nu}$ its dual. Magnetometry measurements \cite{Abel:2017rtm} constrain the strength of this interaction by
\begin{equation}\label{AxionBound}
	\frac{c_2}{\Lambda_2} + \frac{\tilde c_2}{M_{\rm P}} \lesssim M_{\rm P}^{-1}
	\qquad {\rm if} \qquad 
	m_a \lesssim 5\cdot 10^{-21}\, {\rm eV}.
\end{equation}
Therefore, any dark matter model containing scalar axion-like fields of masses $m_a\lesssim 10^{-21} \, {\rm eV}$ is excluded. The result for this particular interaction are summarized in figure \ref{AxionPlot}, which can be compared to figure 4 in ref. \cite{Abel:2017rtm} and figure 31.5 in ref.~\cite{Stadnik:2018sas}. Note that this bound assumes that all of dark matter is described by the axion-like-particle $a$. It is possible to relax this bound if dark matter has multiple components.

On the other hand, for interactions of the form
\begin{equation}\label{AxionLin}
	O_{3,{\rm QG}} = \left (\frac{c_3}{\Lambda_3}+\frac{\tilde c_3}{M_{\rm P}} \right ) \, a\, F_{\mu\nu}\tilde{F}^{\mu\nu},
\end{equation}
with $\tilde c_3\sim {\cal O}(1)$, the bounds are much weaker\footnote{cf. Figure 31.4 in ref.~\cite{Stadnik:2018sas}.}. Therefore, there is still a large parameter space to explore. However, the bound \eqref{AxionBound} excludes axion like particles with masses below $10^{-21} \, {\rm eV}$, because of the universality of gravity: one cannot have the interaction $a  F_{\mu\nu}\tilde{F}^{\mu\nu}$ without  the interaction $a G_{\mu\nu}\tilde{G}^{\mu\nu}$.

\begin{figure}[h]
	\centering
	\includegraphics[width=8cm]{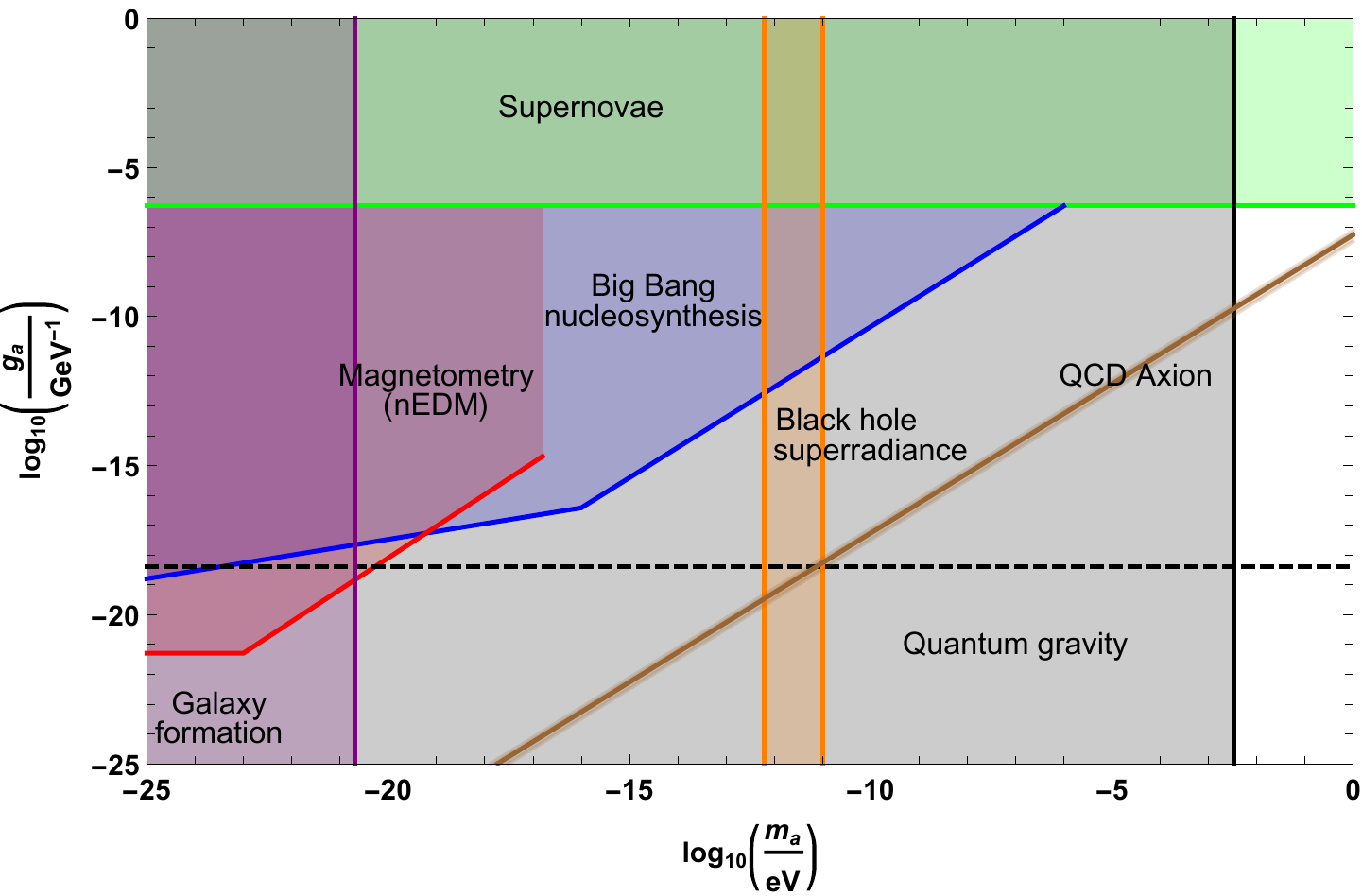}
	\caption{Parity violating quantum gravity. Limits on the linear axion interaction $g_a=c_3/\Lambda_3$ as a function of the mass of the axion $m_a$. Green: limits from supernovae measurements \cite{Graham:2013gfa}. Blue: limits from the big bang nucleosynthesis \cite{Blum:2014vsa,Stadnik:2015uka,Stadnik:2015kia,Stadnik:2015upa,Stadnik:2017mid}. Red: limits from magnetometry experiments \cite{Abel:2017rtm,Stadnik:2018sas}. Purple: limits from galaxy formation, quasar lensing  and stellar streams \cite{Bozek:2014uqa,Schive:2015kza,Corasaniti:2016epp,Irsic:2017yje,Nori:2018pka,Schutz:2020jox}. Orange: limits from the superradiance instability of black holes \cite{Cardoso:2018tly}, however note that these bounds can be avoided, if the self-interaction of the axion-like particle is sufficiently strong \cite{Arvanitaki:2014wva}.  Brown: predicted value of the QCD axion \cite{Pospelov:1999mv,Tanabashi:2018oca}. Black: axion masses below $m_a\lesssim 3 \times 10^{-3}\, {\rm eV}$ are excluded by parity violating quantum gravity as discussed in this paper. Dashed black line: reduced Planck scale.}
	\label{AxionPVPlot}
\end{figure}

Furthermore, there is no reason why parity symmetry would be preserved by quantum gravitational interactions, see e.g. \cite{Holman:1992us,Barr:1992qq}. Indeed, it is not a gauge interaction. In this case, the operators 
\begin{equation}\label{AxionParViol}
	O_4= \frac{\tilde c_4}{M_{\rm P}}\, a\, G_{\mu\nu}G^{\mu\nu},
\end{equation}
and
\begin{equation}\label{AxionParViol}
	O_5 = \frac{\tilde c_5}{M_{\rm P}}\, a\, F_{\mu\nu}F^{\mu\nu},
\end{equation}
which are parity violating will be generated.  As before we expect $\tilde c_4\sim {\cal O}(1)$ and $\tilde c_5\sim {\cal O}(1)$. These operators lead to a Yukawa-type interaction and thus to a fifth force.  Therefore, if quantum gravity violates parity, axion-like-particle with masses $m_a\lesssim 3 \times 10^{-3}\, {\rm eV}$ are excluded. As shown in figure  \ref{AxionPVPlot}, this reduces the parameter space for axion models massively.

\begin{figure}[H]
	\centering
	\includegraphics[width=8cm]{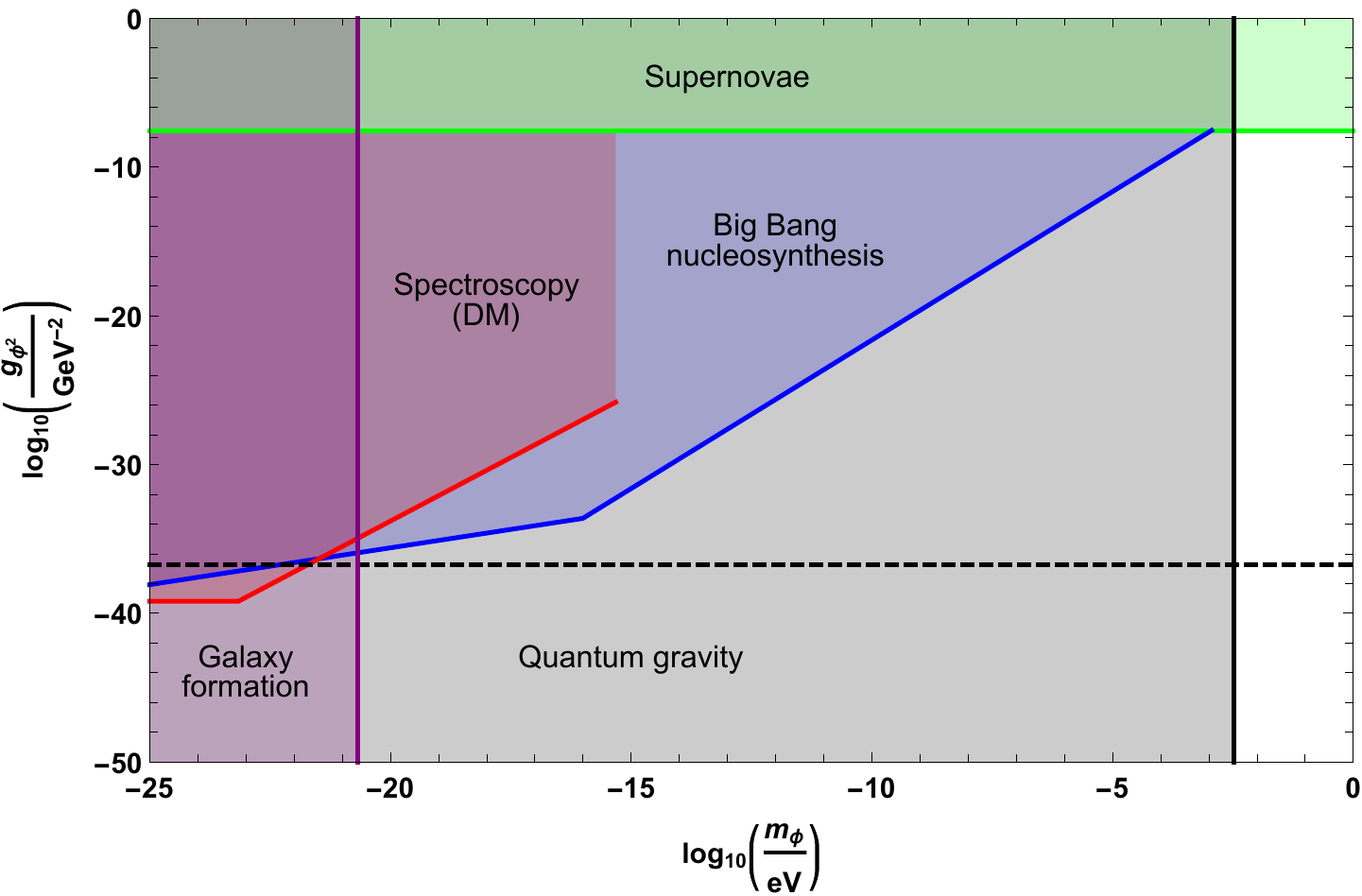}
	\caption{Limits on the quadratic scalar interaction $g_{\phi^2}=c_6/\Lambda_6$ as a function of the mass of the scalar $m_\phi$. Green: limits from supernovae measurements \cite{Olive:2007aj}. Blue: limits from the big bang nucleosynthesis \cite{Stadnik:2015kia}. Red: limits from atomic spectroscopy \cite{Stadnik:2015kia,Stadnik:2016zkf,Stadnik:2018sas}. Purple: limits from galaxy formation, quasar lensing  and stellar streams \cite{Bozek:2014uqa,Schive:2015kza,Corasaniti:2016epp,Irsic:2017yje,Nori:2018pka,Schutz:2020jox}. Black: limits from quantum gravity as discussed in this paper. Dashed black line: reduced Planck scale.}
	\label{Phi2Plot}
\end{figure}

Another possible interaction of a spinless dark matter boson coupling to the standard model is a dimension 6 interaction of the form
\begin{equation}\label{QuadCoup}
	O_{6,{\rm QG}} =\left ( \frac{c_6}{\Lambda_6^2}+\frac{\tilde c_6}{M_{\rm P}^2}\right)\, \phi^2\, F_{\mu\nu}F^{\mu\nu},
\end{equation}
which does not distinguish between scalars and pseudoscalars, as parity is automatically conserved. Again we have $\tilde c_6\sim {\cal O}(1)$. Atomic spectroscopy measurements \cite{Stadnik:2015kia,Stadnik:2016zkf} constrain the strength of this interaction by
\begin{equation}
	\frac{c_6}{\Lambda_6^2} +\frac{\tilde c_6}{M_{\rm P}^{2}} \lesssim M_{\rm P}^{-2} \qquad {\rm if} \qquad  m_\phi \lesssim 2\cdot 10^{-22} \, {\rm eV}.
\end{equation}
Therefore, any dark matter model containing scalar dark matter fields of masses $m_\phi\lesssim 10^{-22}\, {\rm eV}$ that couple to the standard model in this way are excluded. Note that bounds from galaxy formation, quasar lensing  and stellar streams are slightly more stringent and lead to $m_\phi\lesssim 10^{-21}\, {\rm eV}$ but they have a larger uncertainty. Quantum gravity will however also generate operators of the type $M_{\rm P}^{-1} \phi F_{\mu\nu}F^{\mu\nu}$ and $M_{\rm P}^{-1} \phi F_{\mu\nu}\tilde F^{\mu\nu}$ even if these operators are not introduced in the interaction Lagrangian and we can thus rule out masses below $3 \times 10^{-3}\, {\rm eV}$. In the case of axions, this bound applies if parity is violated by quantum gravity which we argued is to be expected. The results are summarized in figure \ref{Phi2Plot}, which can be compared to figure 31.6 in ref.~\cite{Stadnik:2018sas}.

Our results rule out most of the parameter range for ultralight and very weakly coupled singlet scalar dark matter models. It is worth mentioning that our bound applies as well to the quintessence type models which are often advocated to generate a cosmological time evolution of fundamental constant. A change of the hyperfine constant within the last Hubble time, implies the existence of a scalar field with a very light mass of the order of the present Hubble scale $H=10^{-33}$ eV \cite{Dvali:2001dd}. This is ruled out because of quantum gravity. If a time variation of the hyperfine constant is observed, we can safely conclude that it is not due to such a scalar field or dark matter. 

Also, it had already been pointed out that the axion is not a valid solution to the strong CP problem of quantum chromodynamics because quantum gravitational effects would destabilize its potential \cite{Holman:1992us,Barr:1992qq}, our results imply that the quantum chromodynamics axion is ruled out for most of its parameter range because of quantum gravity if parity is, as expected, violated by quantum gravitational effects.

Obviously there is a well known mechanism to avoid the bound from the E\"ot-Wash experiment namely the screening mechanism. However, if the masses of light scalar fields were screened by the matter density on Earth thereby increasing their masses on Earth, they would also be heavy for atomic clocks and quantum sensor experiments based on Earth and would thus not lead to the usual signatures mimicking a time variation of fundamental constants. Interestingly, this could be probed by putting atomic clocks or quantum sensor experiments on a satellite where the screening mechanism would be inefficient.

\begin{figure}[H]
	\centering
	\includegraphics[width=8cm]{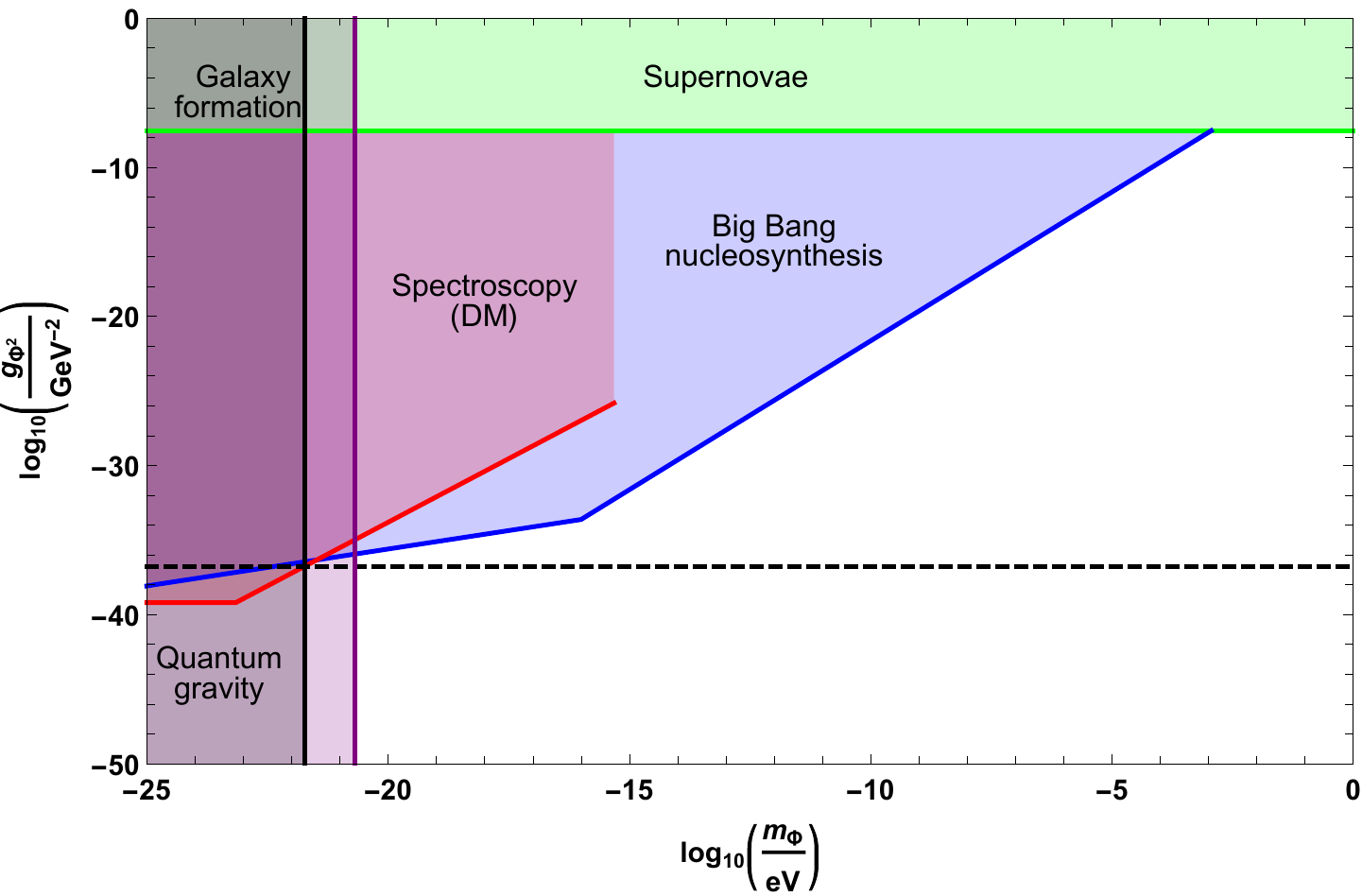}
	\caption{Limits on the quadratic gauged scalar interaction $g_{\Phi^2}=c_7/\Lambda_7$ as a function of the mass of the scalar $m_\Phi$. Green: limits from supernovae measurements \cite{Olive:2007aj}. Blue: limits from the big bang nucleosynthesis \cite{Stadnik:2015kia}. Red: limits from atomic spectroscopy \cite{Stadnik:2015kia,Stadnik:2016zkf,Stadnik:2018sas}. Purple: limits from galaxy formation, quasar lensing  and stellar streams \cite{Bozek:2014uqa,Schive:2015kza,Corasaniti:2016epp,Irsic:2017yje,Nori:2018pka,Schutz:2020jox}. Black: limits from quantum gravity as discussed in this paper. Dashed black line: reduced Planck scale.}
	\label{Phi2PlotGauged}
\end{figure}

While we focussed thus far on scalar and pseudoscalar fields which are singlets under gauge symmetries, it is possible to avoid some of the bounds from quantum gravity discussed above if we consider scalar or pseudoscalar fields that are gauged under some new gauge group, as gauge symmetries are preserved by quantum gravity. In that case, the only relevant operators are dimension 6 ones of the type 
\begin{equation}\label{QuadCoupgauged}
	O_{7,{\rm QG}} =\left ( \frac{c_7}{\Lambda_7^2}+\frac{\tilde c_7}{M_{\rm P}^2}\right)\, \Phi \cdot \Phi \, F_{\mu\nu}F^{\mu\nu},
\end{equation}
where $\Phi$ is a scalar or pseudoscalar field gauged under some new gauge group of the dark matter sector and $ \Phi \cdot \Phi$ is a scalar under that gauge symmetry. We find
\begin{equation}
	\frac{c_7}{\Lambda_7^2} +\frac{\tilde c_7}{M_{\rm P}^{2}} \lesssim M_{\rm P}^{-2} \qquad {\rm if} \qquad  m_\Phi \lesssim 2\cdot 10^{-22} \, {\rm eV}.
\end{equation}
in which case we can only exclude masses $m_\Phi\lesssim 10^{-22}\, {\rm eV}$ for scalar and pseudoscalar fields (or $m_\Phi\lesssim 10^{-21}\, {\rm eV}$  if we use the bound from galaxy formation, quasar lensing  and stellar streams \cite{Bozek:2014uqa,Schive:2015kza,Corasaniti:2016epp,Irsic:2017yje,Nori:2018pka,Schutz:2020jox}). If atomic clocks or quantum sensor experiments were to discover such scalar or pseudoscalar fields, they would not only have discovered dark matter but also proven the existence of a new gauge force in the dark matter sector. The results are summarized in figure \ref{Phi2PlotGauged}. For quintessence fields, the effect would be of order $(\Delta \phi/M_{\rm P})^2$ and thus more suppressed than usually assumed.

Let us finally emphasize that the bounds on quantum gravity shown in figures \ref{PhiPlot}, \ref{AxionPlot}, \ref{AxionPVPlot}, \ref{Phi2Plot} and \ref{Phi2PlotGauged} carry a small theoretical uncertainty, as the Wilson coefficients are not exactly known. We argued that we know the scale of quantum gravity and that it can be calculated given the number of fields introduced in the model. While the scale of quantum gravity incorporates any suppression for the operators generated by quantum gravity, it is conceivable that the Wilson coefficients could take values between $10^{-1}$ and 10. Smaller than unity Wilson coefficients could still decrease the bounds by about a factor of $10$, which would bring the bound from $g=4\times10^{-19} {\rm GeV}^{-1}$ to $g=4\times10^{-20} {\rm GeV}^{-1}$ in figures \ref{PhiPlot}, \ref{AxionPlot} and \ref{AxionPVPlot}, and from $g=2\times10^{-37} {\rm GeV}^{-1}$ to $g=2\times 10^{-39} {\rm GeV}^{-1}$ in figure \ref{Phi2Plot}. If the Wilson coefficients were order $10^{-1}$, we could only exclude masses below $1\times 10^{-4} {\rm eV}$.

Moreover, the bounds derived from spectroscopy experiments (red lines) and from models of galaxy formation, are based on the assumption that the scalar field accounts for the total observed local dark matter density $\rho=0.4 {\rm GeV/cm^3}$. Multicomponent dark matter models would loosen the bounds shown in figures \ref{PhiPlot}, \ref{AxionPlot},  \ref{AxionPVPlot} and \ref{Phi2Plot}.

\section{Conclusions}

In this paper we have considered models of dark matter with ultra-light scalar or pseudoscalar fields which have received a lot of attention as they could be discovered with tabletop experiments looking for dark matter using modern quantum sensors or atomic clocks. These particles are usually assumed to be extremely light and very weakly coupled to the particles of the standard model. 

We have argued that quantum gravity will induce interactions between scalar or pseudoscalar dark matter particles and those of the standard model. These quantum gravitational interactions often dominate over the strength of the interaction posited in these models. We have shown that these quantum gravitational interactions are of the fifth force type for scalar dark matter and also for pseudoscalar dark matter if quantum gravity violates parity symmetry. Such interactions are constrained by torsion pendulum experiments such as the E\"{o}t-Wash experiment. Scalar dark matter must be heavier than $3\times 10^{-3}$ eV and the same bound applies to pseudoscalar particles assuming that quantum gravity violates parity symmetry. If quantum gravity does not violate parity, pseudoscalar particles are only constrained to have masses larger than $10^{-21} \, {\rm eV}$. We stress that these bounds are universal and applicable to any scalar dark matter models including for example models for fuzzy dark matter (see e.g. \cite{Hu:2000ke,Marsh:2013ywa,Schive:2014dra}) which have recently received a lot of attention.

While singlet scalar or pseudoscalar fields are constrained to be heavier than $3\times 10^{-3}$ eV, gauged fields could be much lighter. They could be as light as $m_\Phi \sim 10^{-22}\, {\rm eV}$ and thus very much relevant to current experiments using atomic clocks or quantum sensors.  A positive signal would not only be potentially the sign of dark matter but also a sign that the dark matter sector is very rich and contains new forces. Another way to look at our results is that very low energy tabletop experiments such as atomic clocks and other experiments based on quantum sensors are directly probing quantum gravitational effects.

\section*{Acknowledgments}
The authors would like to thank Yevgeny Stadnik for very helpful discussions and comments on the draft of this paper.
The work of X.C.~is supported in part  by the Science and Technology Facilities Council (grant number ST/P000819/1).
The work of F.K.~is supported by a doctoral studentship of the Science and Technology Facilities Council.

\end{document}